# Waveguide-Integrated Two-Dimensional Material Photodetectors in Thin-Film Lithium Niobate

Sha Zhu, Yiwen Zhang, Yi Ren, Yongji Wang, Kunpeng Zhai, Hanke Feng, Ya Jin, Zezhou Lin, Jiaxue Feng, Siyuan Li, Qi Yang, Ning Hua Zhu, Edwin Yue-Bun Pun*, and Cheng Wang*

**Abstract:**

Thin-film lithium niobate on insulator (LNOI) is a promising platform for optical communications, microwave photonics, and quantum technologies. While many high-performance devices like electro-optic modulators and frequency comb sources have been achieved on LNOI platform, it remains challenging to realize photodetectors (PDs) on LNOI platform using simple and low-cost fabrication techniques. Two-dimensional (2D) materials are excellent candidates to achieve photodetection since they feature strong light-matter interaction, excellent mechanical flexibility, and potential large-scale complementary metal-oxide-semiconductor-compatible fabrication. In this work, we propose to address this demand using an LNOI-2D material platform and demonstrate two types of high-performance LNOI waveguide-integrated 2D material PDs, namely graphene and Tellurium (Te). Specifically, the LNOI-graphene PD features broadband operations at telecom and visible wavelengths, high normalized photocurrent-to-dark current ratios up to $3\times10^6$ W$^{-1}$, and large 3-dB photoelectric bandwidths of over 40 GHz, simultaneously. The LNOI-Te PD on the other hand provides an ultrahigh responsivity of 7 A/W under 0.5 V bias for telecom optical signals while supporting GHz frequency responses. Our results show that the versatile properties of 2D materials and their excellent compatibility with LNOI waveguides could provide important low-cost solutions for system operating point monitoring and high-speed photoelectric conversion in future LN photonic integrated circuits.

## Introduction

Photonic integration plays an important role in future ultra-high-speed optical communications, microwave signal processing, as well as quantum computation and communications. Lithium niobate (LN) is one of the most promising optical materials for future integrated photonics, exhibiting a high electro-optic coefficient ($r_{33}$ = 27 pm/V), relatively large refractive indices ($n_o$ = 2.21 and $n_e$ = 2.14), a wide transparency window, and low-optic losses[1]. Traditional LN devices formed by titanium in-diffusion or proton exchange typically feature weak optical confinement due to the low refractive index contrast (~0.02), and as a result compromised device performance. In comparison, the recently emerged thin-film LN platform benefits from a much larger index contrast and has enabled a range of compact and high-performance integrated photonic components, including electro-optic modulators[2], frequency combs[3], delay lines[4], lasers[5], optical filters[6], acousto-optic modulators[7] and so on. More importantly, the commercial availability of high-quality LNOI wafers at scales up to 6 inches and the recent demonstration of wafer-scale device manufacturing processes have further boosted the competitiveness of the LNOI platform in terms of scalability, cost-effectiveness, and commercial readiness[8].

Photodetectors (PDs) are indispensable components for recovering transmitted signals in optical communications, photo-electric conversion in microwave photonics, and monitoring system bias points in large-scale photonic networks[9]. Nevertheless, integrating PDs on the LNOI platform has remained a substantial challenge due to LN's intrinsically high transparency in both visible and infrared wavelengths, as well as lattice mismatch issues that prevent the epitaxial growth of III-V materials on LN. Attempts have been made to heterogeneously integrate III-V PDs with LNOI

S. Zhu (zhusha@bjut.edu.cn), J. Feng (fengjiax@emails.bjut.edu.cn): College of Microelectronics, Faculty of Information Technology, Beijing University of Technology, Beijing, 100124, China

S. Zhu (shazhu@cityu.edu.hk), Y. Zhang (yzhang2544-c@my.cityu.edu.hk), H. Feng (hankefeng2-c@my.cityu.edu.hk), E. Y. B. Pun (eeeybpun@cityu.edu.hk), C. Wang (cwang257@cityu.edu.hk): Department of Electrical Engineering and State Key Laboratory of Terahertz and Millimeter Waves, City University of Hong Kong, Kowloon, Hong Kong, China

Y. Ren (yiren26-c@my.cityu.edu.hk), Y. Wang (yongjwang5-c@my.cityu.edu.hk), S. Li (siyuan.li@my.cityu.edu.hk), Q. Yang (qyang62-c@my.cityu.edu.hk): Department of Chemistry, City University of Hong Kong, Hong Kong, China

K. Zhai (kpzhai@semi.ac.cn), Y. Jin (jinya@semi.ac.cn), N. H. Zhu (nhzhu@semi.ac.cn): State Key Laboratory on Integrated Optoelectronics, Institute of Semiconductors, Chinese Academy of Sciences, Beijing, 100083, China

Z. Lin (21037392r@connect.polyu.hk): Department of Applied Physics and Research Institute for Smart Energy, The Hong Kong Polytechnic University, Hong Kong, China

Correspondence: E. Y. B. Pun and C. Wang

These authors contributed equally: S. Zhu, Y. Zhang, Y. Ren, Y. Wang

This project was supported in part by National Natural Science Foundation of China (61922092); Research Grants Council, University Grants Committee (CityU 11204820, CityU 11212721, N_CityU113/20); Croucher Foundation (9509005); CityU Strategic Grant 7005458; High-end Talent Team Construction Plan of Beijing University of Technology.

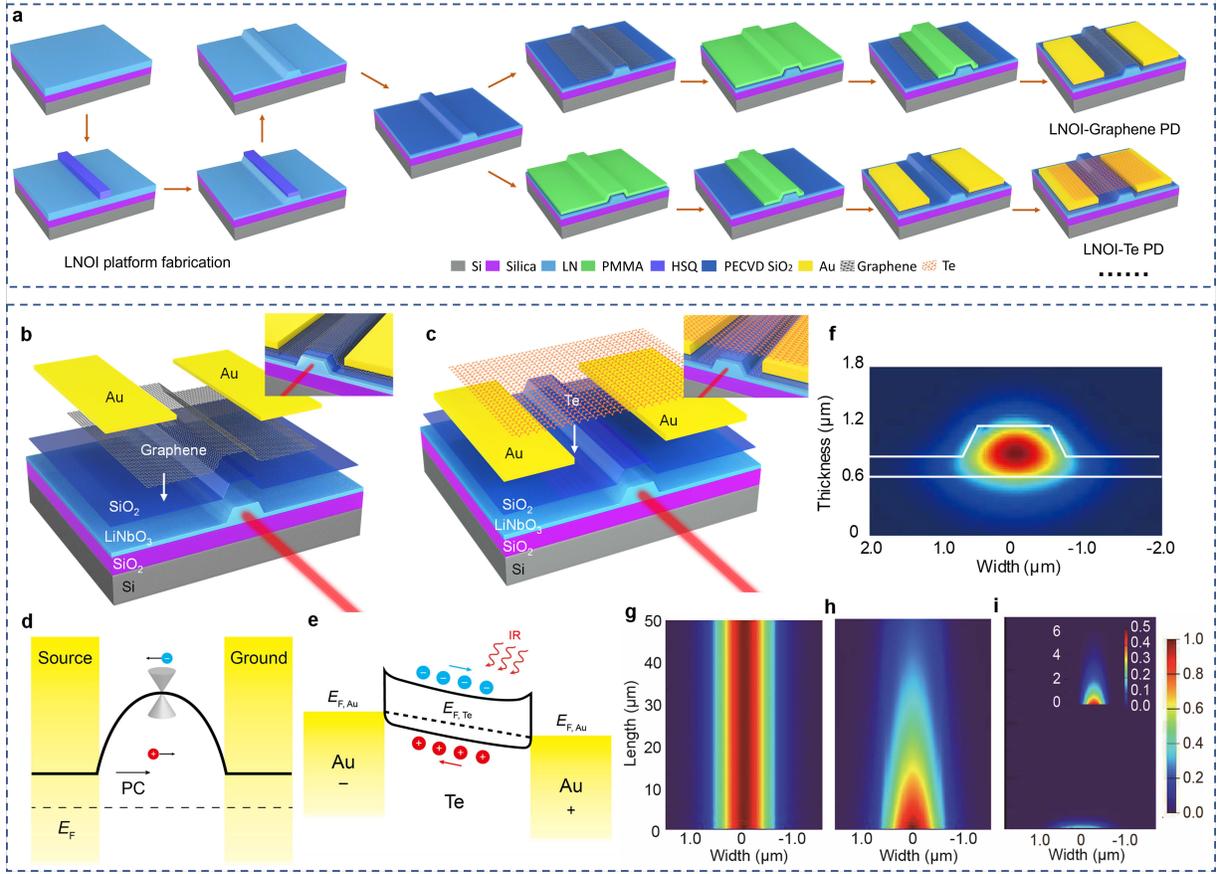

Fig. 1 a) Fabrication flowchart of the waveguide-integrated 2D material PDs in LNOI platform. 3D schematic and zoom-in view of the b) LNOI-graphene and c) LNOI-Te PDs. Energy band diagrams of the d) LNOI-graphene PD and e) LNOI-Te PDs. $E_F$: Fermi level, PC: photocurrent, WG: waveguide, IR: infrared, $E_{F,Au}$ and $E_{F,Te}$: Fermi levels of Au and Te, respectively. f) Simulated optical eigenmode profile (electric field intensity) of the LNOI waveguide. Simulated electric field intensity evolution along the LNOI waveguide g) without 2D material, with 10 layers of h) graphene and i) Te.

devices via die bonding[10]. Such processes, however, require expensive and complicated alignment and bonding tools to achieve good yield at large scales. More scalable and cost-effective integration of PDs on the LNOI platform could be realized using deposited amorphous silicon, which however works in the visible wavelength range only[11]. Two-dimensional (2D) materials could become an excellent alternative to achieve PDs on the LNOI platform in a simple and widely compatible way. Stacked via van der Waals forces, the atomically thin layers of 2D materials feature strong light-matter interactions, superior mechanical flexibility, and good chemical stability without surface dangling bonds[12], leading to high responsivities, good long-term operational stability, and excellent compatibility with different photonic materials and structures. Previously demonstrations of LNOI-2D material integration however have been limited to etch-less waveguides with flat LN surfaces and to graphene only, where the bound-state-in-the-continuum (BIC) modes only support low-loss propagation for certain waveguide dimensions [13].

Here, we demonstrate a general and versatile LNOI-2D material PD platform by integrating two types of 2D materials with monolithically fabricated LNOI waveguides. Specifically, graphene PDs provide broad electrical and optical operation bandwidths, owing to its ultra-high carrier mobility, broadband optical absorption, and tunable Fermi level[14], whereas tellurium (Te) PDs feature extremely high responsivities due to the carrier avalanche and photogating effect[15]. Furthermore, both graphene and Te feature good long-term stability in air and potentially allow wafer-scale integration with the LNOI platform, i.e. by transferring chemical vapor deposited graphene films and thermal evaporation of Te films[16]. We first demonstrate a broadband LNOI-graphene PD operating at both telecommunication and visible bands, with a low dark current of 5 nA and a high normalized photocurrent-to-dark current ratio (NPDR) above $3\times10^6$ W$^{-1}$ at 1550 nm. More importantly, our

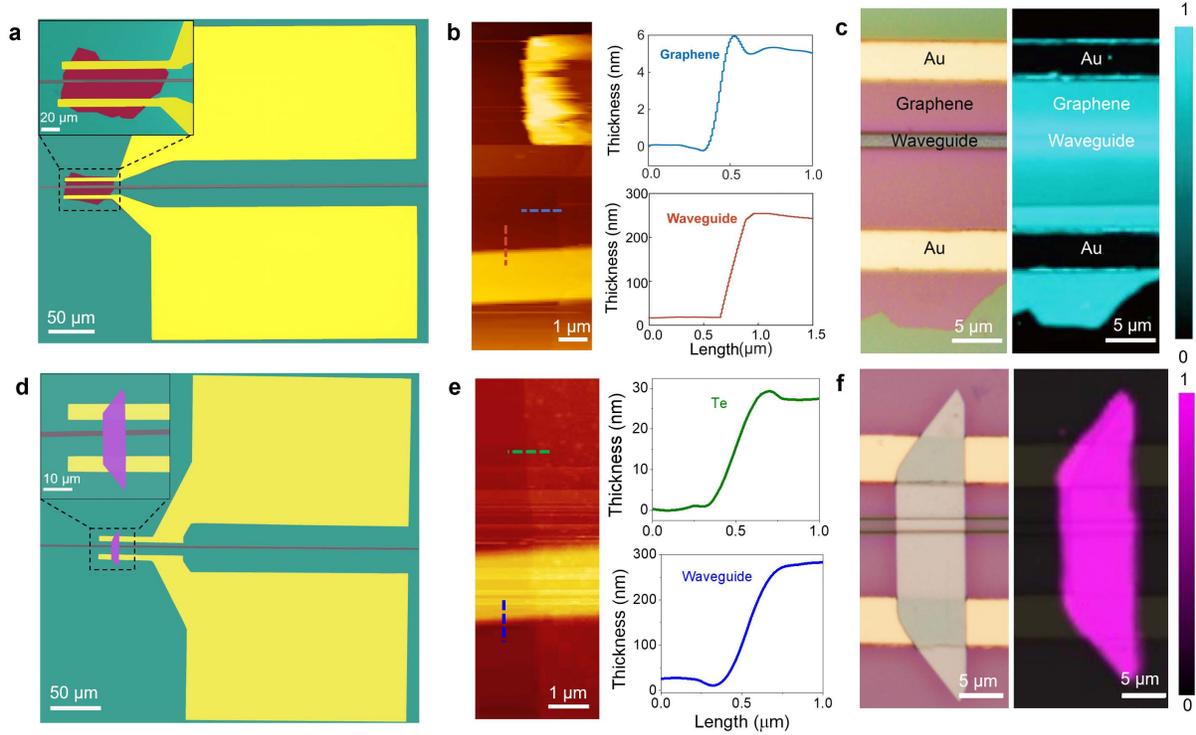

Fig. 2 a) False-colored scanning electron microscope (SEM) image with zoom-in view, b) atomic force microscope (AFM) image, c) optical image (left) and Raman mapping image (right) of the fabricated LNOI-graphene PD. d) False-colored SEM image with zoom-in view, e) AFM image, f) optical image (left) and Raman mapping image (right) of the fabricated LNOI-Te PD.

LNOI-graphene PD features near-flat photoelectric responses at frequencies up to 40 GHz, limited by testing equipment. We then demonstrate an LNOI-Te PD with an ultrahigh responsivity of 7 A/W under a 0.5 V bias at 1550 nm and a 3 dB bandwidth of 2 GHz. The versatile performances provided by our LNOI-2D material platform could serve as important building blocks for various photodetection needs in future LNOI photonic integrated circuits.

Figures 1(a)-1(c) depict the fabrication flow and 3D schematics of the two proposed LNOI-2D material PDs. For 2D materials with excellent mechanical flexibility and toughness such as graphene, it can directly cover the surface of LNOI waveguide without any wrinkles and rupture, as shown in Fig. 1(b). The electrodes are then deposited on top of the transferred graphene. For other 2D materials like Te, we deposit the electrodes first to reduce the height difference between the waveguide and the surrounding plateau, so that the 2D material can be transferred with less stress and potential cracks, as shown in Fig. 1(c). On the surface of LN, a thin $SiO_2$ gap layer is deposited for electrical isolation and contamination control.

Figure 1(d) shows the energy band diagram of the LNOI-graphene PD. Due to the different doping levels between the graphene under gold electrodes (p-doped[17]) and the graphene channel (p$^+$-doped induced by surrounding air), the interfaces of the two electrodes exhibit obvious band bending, reverse potential gradients and also a built-in electric field. This enables the operation of our PD in a self-powered state without the need for a DC bias voltage when we design the pair of electrodes to be asymmetric [inset of Fig. 1(b)][18]. The potential gradient induced by the gold electrode closer to the waveguide substantially overlaps with the optical field and could efficiently separate the generated photocarriers even at zero bias, while the gold electrode farther from the waveguide moves electrons (or holes) unidirectionally to avoid the cancellation of the overall photocurrent[14b, 19]. We use finite-difference time-domain (FDTD, Ansys Lumerical) simulation to verify that the optical signal could be efficiently absorbed by our multilayer graphene. Figure 1(f) shows the cross-sectional optical eigenmode profile of the LN waveguide for $TE_0$ mode, which aligns with the largest electro-optic coefficient ($r_{33}$) in our $x$-cut LNOI material and is the most relevant for majority of applications. Figures 1(g) and 1(h) show the simulated optical transmission performance along the LN waveguide without and with the graphene layer, indicating a strong absorption of optical signals within a few tens of microns. We have also calculated the light absorption for different numbers of graphene layers, as shown in Fig. S1. For single-layer graphene, the absorption coefficient is only 0.014 dB/μm, while for 10-layer graphene used in this work, the

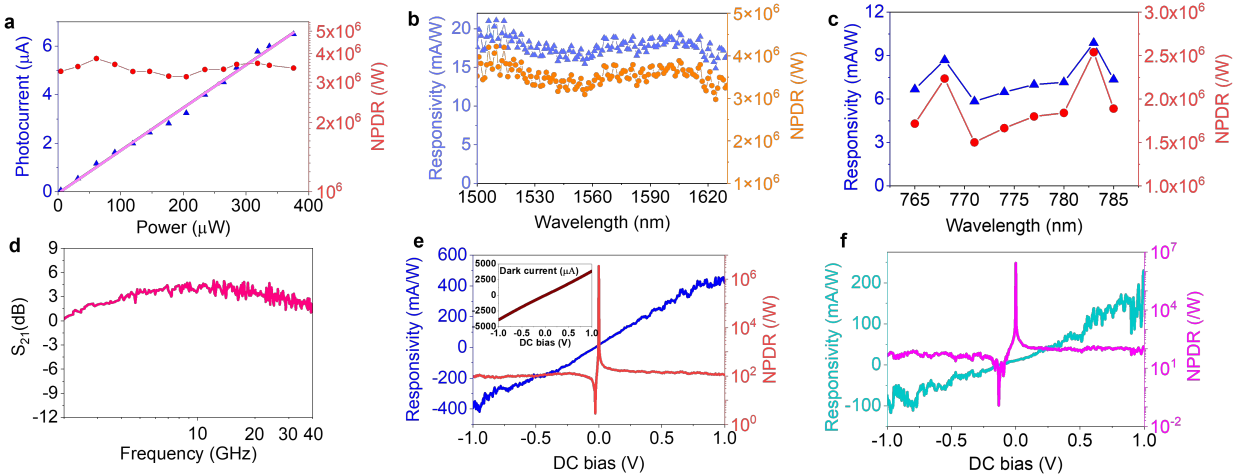

Fig. 3 a) Measured photocurrent and normalized photocurrent-to-dark-current ratios (NPDR) of the LNOI-graphene PD versus input optical power at 1550 nm under zero bias. Measured responsivity and NPDR at b) telecom and c) visible bands under zero bias. d) Measured photoelectric $S_{21}$ response at 1550 nm. Measured responsivity and NPDR with respect to DC bias at e) 1550 nm (Inset: dark current versus DC bias) and f) 783 nm wavelength.

absorption coefficient can reach up to 0.14 dB/μm. More detailed simulated electric field intensity evolution for different numbers of layers can be found in Fig. S2.

For the LNOI-Te PD, its energy band diagram under a finite electrical bias is shown in Fig. 1(e). The finite bandgap (~0.27 eV) of Te flake allows the absorption of infrared light in the LNOI waveguide and the generation of photocurrents via photoconductive effect[15d]. The source-drain electrodes are separated from the light absorption region by a distance larger than the width of the depletion region. As a result, the LNOI-Te PD exhibits a relatively small photocurrent at 0 V bias, ruling out the possibility that the produced photocurrent is attributed to photovoltaic or photo-thermoelectric effects. Instead, the photogenerated charge carriers are moved directionally by the electrostatic force when a DC bias voltage is applied and finally collected by the gold electrodes. Figure 1(h) shows the simulated optical transmission performance of the LNOI-Te PD with a large optical absorption of 3.42 dB/μm (Fig. S3). The details of device fabrication and numerical simulation can be found in Methods.

Next, we perform a detailed material characterization of the fabricated LNOI-2D material devices, which is key to the performances of the final PDs (Fig. 2). For the LNOI-graphene PD, Fig. 2(a) shows its false-colored scanning electron microscope (SEM) image, where the multilayer graphene of about 56 μm in length is tightly attached to the LN waveguide without wrinkles or ruptures, ensuring good conductivity and efficient optical absorption. To further reveal the height profile of the graphene and the LNOI waveguide, atomic force microscope (AFM) measurement is carried out, which reveals that the thicknesses of multilayer graphene and LN waveguide are about 5 nm and 250 nm, respectively [Fig. 2(b)]. The microscopic atomic structure of the multilayer graphene is further characterized by transmission electron microscopy (TEM), where selected area electron diffraction (SAED) confirms its single crystalline nature (Fig. S4)[20]. Raman spectra collected at the transferred multilayer graphene area also show clear characteristic Raman peaks at 1581 cm$^{-1}$ (G) and 2717 cm$^{-1}$ (2D)[21], with an intensity ratio of 2D/G less than 1, which indicates that the attached material is indeed multilayer graphene (Fig. S5)[22]. We perform Raman intensity mapping of the PD at 1581 cm$^{-1}$ [Fig. 2(c)], confirming that the graphene channel still maintains good integrity and uniformity without cracks and wrinkles after transfer, which lays a solid foundation for the excellent PD performance to be discussed next.

Similarly, we perform detailed material characterizations of the LNOI-Te PD to confirm its single-crystalline properties and structural integrity after transfer. Figures 2(d) and (f) show the false-colored SEM and Raman intensity mapping of the LNOI-Te PD, confirming the 8 μm wide Te flake covers the LNOI waveguide and electrodes evenly without wrinkles. AFM measurement [Fig. 2(e)] reveals a Te layer height of 27 nm, which provides a bandgap (~0.27 eV) to support light absorption at the telecom band[15a, 23]. TEM and SAED measurements verify the single crystalline nature of Te (Fig. S6)[15a], whereas the Raman peaks observed at 92 cm$^{-1}$ ($E_1$), 121 cm$^{-1}$ ($A_1$), and 140 cm$^{-1}$ ($E_2$, used for Raman mapping in Fig. 2(f)) confirm that the material covering the LN waveguide is indeed Te [15a].

We then show that our LNOI-2D material PDs could provide excellent complementary performances suitable for different application scenarios in future LNOI photonic circuits. In particular, the LNOI-graphene PD provides broadband high-speed photoelectric responses from near-visible to telecom bands with low dark currents. Benefiting from the built-in electric field induced by the asymmetric electrodes, substantial photocurrent ($I_p$) could be generated even at zero bias, allowing for a low dark current ($I_d$) of 5 nA, defined as the current collected at zero input optical power. The measured photocurrent linearly increases with increasing input optical power ($P_{in}$) without saturation at up to 400 µW on-chip power, as shown in Fig. 3(a), leading to a measured responsivity of 17.27 mA/W at 1550 nm under zero bias. The combination of a good responsivity and low dark current leads to a high normalized photocurrent-to-dark current ratio (NPDR) above $3\times10^6$ $W^{-1}$ [Fig. 3(a)], indicating an excellent signal-to-noise performance in our PDs.

Our LNOI-graphene PD could efficiently operate over a broad wavelength range benefiting from the absence of bandgap in graphene. Figure 3(b) and (c) display the measured responsivity and NPDR under zero bias at different telecom (1500-1630 nm) and near-visible (765-785 nm) wavelengths, showing broadband near-flat spectral responses in both bands. Compared with the telecom band, the responsivity at near-visible is slightly lower (~ 8 mA/W) due to a smaller optical mode and consequently a reduced modal overlap with graphene. Still it exhibits a high NPDR of $1.7\times10^6$ $W^{-1}$ thanks to the low dark current. Since graphene has no band gap and our LNOI waveguides support low-loss light transmission over a broad wavelength range, we expect the actual operation bandwidth of our PD to cover the entire visible, near-infrared, and part of the mid-infrared bands (ultimately limited by waveguide cutoff and $SiO_2$ material absorption).

Furthermore, we show that our LNOI-graphene PD could provide a near-flat frequency response up to at least 40 GHz, currently limited by the testing equipment, as the measured photoelectric $S_{21}$ in Fig. 3(d) shows. The small increase in $S_{21}$ at low frequencies could be attributed to an unmatched source impedance in our PD[19b], and could be further optimized by designing the resistance and capacitance of the PD electrode pads. We expect our LNOI-graphene PD to efficiently operate at frequencies much beyond 40 GHz considering the high mobility of graphene and the small active area in our device.

The responsivity of the LNOI-graphene PD can be further increased significantly by applying an external DC bias to induce an additional electrical field in the active region [Fig. 3(e)]. For positive bias voltages, the responsivity can reach up to 419 mA/W at a 1 V bias without observable saturation behavior. When the bias voltage turns negative, the internal electrical field is first canceled and then reversed, leading to a negative responsivity. Under -0.028 V DC bias, the responsivity becomes zero. The larger responsivities at finite bias voltages are accompanied with increased dark currents [inset of Fig. 3(e)], leading to lowered NPDR on the order of $10^2$ $W^{-1}$ [Fig. 3(e)], still comparable with other 2D-material PDs[24]. To achieve high responsivity and NPDR simultaneously, LNOI-Te PDs could be used instead, as we discuss next. Similar trend of responsivity increase at finite bias voltages can also be observed in the near-visible band, reaching 200 mA/W at 1 V and -100 mA/W at -1 V [Fig 3(f)]. Detailed measurement setups and characterization methodologies can be found in "Materials and Methods" and supplementary information.

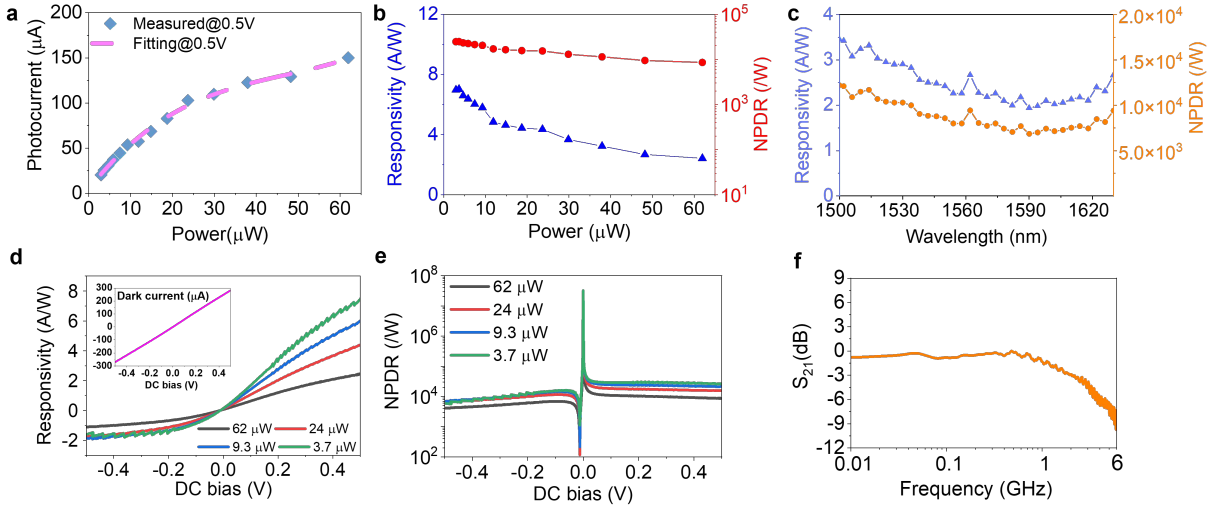

Fig. 4 a) Measured photocurrent as well as b) responsivity and NPDR versus input optical power at 1550 nm under 0.5 V DC bias. c) Measured responsivity and NPDR from 1500 nm to 1630 nm wavelength under 0.5 V DC bias. d) Measured responsivity (Inset: dark current versus DC bias) and e) corresponding NPDR versus DC bias at 1550 nm. f) Measured frequency response of the LNOI-Te PD.

Finally, we show the LNOI-Te PD could offer ultrahigh responsivity and high NPDR at the same time, ideally suited for system monitoring of photonic circuits. As Figs. 4(a) and 4(b) show, the measured responsivity exceeds 7 A/W at a weak input optical power of 3 µW at 1550 nm under a bias of 0.5 V, thanks to the strong light absorption and high photogain of Te material. The photocurrent sees saturation behavior and reduced responsivities at higher optical powers (but remains > 2 A/W at 60 µW) due to the limited trap states in Te, as shown in Figs. 4(a) and 4(b). Figure 4(c) shows the measured responsivity and NPDR from 1500 nm to 1630 nm at an optical power of 50 µW under 0.5 V bias. The responsivity is nearly flat with high NPDRs on the order of $10^4$ $W^{-1}$ throughout the measured telecom band. As illustrated in Figs. 4(d) and 4(e), although the dark current does increase with DC bias (inset) similar to that in graphene PD, the LNOI-Te PD benefits from much higher responsivities and lower absolute dark current values, resulting in the high measured NPDRs. We also measured the high-frequency performance of the fabricated LNOI-Te PD, which shows a 3-dB bandwidth of 2 GHz [Fig. 4(f)], limited by the carrier mobility of Te and channel width of the PD structure. By adding gate bias and narrowing the channel, the bandwidth can be further improved. The excellent capability to sense weak optical signals with moderate photoelectric bandwidth make our LNOI-Te PDs ideal for monitoring the variation of optical power in future large-scale LNOI photonic integrated circuits, e.g. DC bias drift of modulators, without the need to tap a substantial power out of the network.

In summary, we fill the photodetection gap on the LNOI platform by demonstrating two proof-of-concept waveguide-integrated LNOI-2D material PD platforms with complementary performance edges. The broadband LNOI-graphene PDs, with a high NPDR of $3\times10^6$ $W^{-1}$, a decent responsivity of 17.27 mA/W at zero bias, and a high photoelectric bandwidth in excess of 40 GHz, could provide high-speed optical receiving elements in future LNOI-based optical communications and microwave photonic systems. The LNOI-Te PDs with ultrahigh responsivities up to 7 A/W and bandwidths over 2 GHz could serve as highly sensitive operating point monitoring components in future LNOI photonic networks for optical computation and signal processing applications. Importantly, our LNOI-2D material platform could potentially be co-fabricated on wafer scales at low cost, by integrating CVD-grown graphene and/or evaporated Te thin films on LNOI wafers processed using stepper lithography processes. Further integrating LNOI waveguides with other 2D materials like transition metal dichalcogenides ($MoS_2$, $MoTe_2$, $WSe_2$) and noble metal dichalcogenides ($PtSe_2$, $PdSe_2$) could provide even more versatile photodetection properties as well as other active photonic functionalities in the LNOI platform to serve the wide variety of photonic system applications in the future.

## Materials and methods

Numerical Simulation.

Ansys Lumerical Finite-Difference-Time-Domain (FDTD) software is used to simulate the cross-section eigenmode profiles and electric field intensity evolutions in our LNOI-2D material PDs. A surface conductivity model is used for the multi-layer graphene, where a scattering rate of 15 meV, a chemical potential of 0.11 eV and a temperature of 300 K are employed[25]. The complex refractive index of Te is $n_{Te}$ = 4.579 + 0.161i at 1550 nm. The thicknesses are set according to the AFM measurement results[26]. The absorptance is given by $\eta=1-10^{-0.1\alpha L}$, where $L$ is the propagation distance, $\alpha$ is the mode absorption coefficient in dB/µm.

Device Fabrication.

The flowchart of the fabrication process of the LNOI-graphene PD is shown in Fig. 1(a). The LN waveguides are fabricated on a commercial $x$-cut LNOI wafer (NANOLN) with a 500-nm-thick thin-film LN, a 2 µm buried silica layer, and a 500 µm silicon substrate. We define the optical waveguide using hydrogen silsesquioxane (HSQ) by electron-beam lithography (EBL) and transfer the patterns using an optimized argon plasma-based reactive ion etching (RIE) process to dry etch a rib LN waveguide with a top width of 1.2 µm, a rib height of 250 nm, and a tilted angle of 45°. Then, a 10 nm $SiO_2$ layer is deposited using plasma-enhanced chemical vapor deposition (PECVD) as a protective and insulation layer. Next, for the LNOI-graphene PD, an exfoliated multilayer graphene is transferred to the LN waveguide. An asymmetric electrode pair is defined by EBL and patterned on the LNOI waveguides by thermal evaporation and lift-off. For the LNOI-Te PD, the electrode is first fabricated and then the Te is transferred to cover the LN waveguide and electrodes. Detail synthesis of Te nanoflakes is shown in supplementary information.

Materials characterizations.

The device structures and the morphologies of the materials are directly observed and recorded by an optical microscope (Nikon LV150N, Japan). SEM images are acquired using Hitachi TM4000. The AFM morphology of the PD is characterized by utilizing a Hitachi AFM 5100N under air atmosphere. The Raman spectra and mapping profile are obtained from a Reinshaw inVia Raman microscope equipped with a 532 nm laser. The diameter of the laser spot is approximately 1 µm. TEM measurement is performed using Philips Tecnai 12 BioTWIN operated at 120 kV.

Device characterizations

The DC and high frequency responses of the PDs are measured based on the setups in Fig. S8. A continuous-wave optical carrier emitted from a laser diode (Santec TSL-550 for telecom, Toptica photonics DL PRO 780S for visible) is sent to a fiber polarization controller (FPC) to accurately control the polarization state, and then coupled into an on-chip LN waveguide. The static responses of the PDs are measured by a precision source/measure unit (Keysight B2902A) at different input optical powers. The input power is calibrated by subtracting the coupling loss between lensed fiber and LNOI chip. Using a spot size converter (SSC) at the input edge of the LNOI-graphene PD could substantially reduce the coupling loss [27]. First, the dark currents at different DC biases are measured without any optical signal input. Afterwards, optical signal is coupled into the PD to measure the photocurrent under various input power intensities, optical wavelengths and bias voltages. Finally, for the high-frequency response of the PD, the optical signal is modulated by a commercial Mach-Zehnder modulator (EOSPACE 40+ Gb/s modulator) driven by microwave signals from a VNA (E5080B Vector Network Analyzer 100 kHz-53 GHz). The modulated optical signal is then sent to the chip and detected by our PD. The recovered electrical signal is collected by a high-frequency RF probe and sent back to the VNA to obtain the $S_{21}$ response.

**Associated content**

Supporting Information

The Supporting Information is available free of charge at XXX.

Simulated optical absorption results of the LNOI-2D material PDs. TEM characterizations and Raman spectra of the 2D material for the LNOI-2D PDs. Experiment setups for the direct current and high-frequency response of the LNOI-2D material PD. Synthesis of Te nanoflakes Flowchart of the fabrication process of the LNOI-2D PD.

**Acknowledgements**

We thank to Prof. Jeff Ou for the use of near-visible laser, Dr. Wing-Han Wong and Dr. Keeson Shum for their help in device fabrication and measurement.

**Conflict of Interest**

The authors declare no competing interests

**Keywords**

Photodetector, thin-film lithium niobate on insulator, 2D materials, graphene, tellurium

# Supplementary Information

# Waveguide-Integrated Two-Dimensional Material Photodetectors in Thin-Film Lithium Niobate

Sha Zhu, Yiwen Zhang, Yi Ren, Yongji Wang, Kunpeng Zhai, Hanke Feng, Ya Jin, Zezhou Lin, Jiaxue Feng, Siyuan Li, Qi Yang, Ning Hua Zhu, Edwin Yue-Bun Pun, and Cheng Wang

S. Zhu (zhusha@bjut.edu.cn), J. Xue (fengjiax@emails.bjut.edu.cn): College of Microelectronics, Faculty of Information Technology, Beijing University of Technology, Beijing, 100124, China

S. Zhu (shazhu@cityu.edu.hk), Y. Zhang (yzhang2544-c@my.cityu.edu.hk), H. Feng (hankefeng2-c@my.cityu.edu.hk), E. Y. B. Pun (eeeybpun@cityu.edu.hk), C. Wang (cwang257@cityu.edu.hk): Department of Electrical Engineering and State Key Laboratory of Terahertz and Millimeter Waves, City University of Hong Kong, Kowloon, Hong Kong, China

Y. Ren (yiren26-c@my.cityu.edu.hk), Y. Wang (yongjwang5-c@my.cityu.edu.hk), S. Li (siyuan.li@my.cityu.edu.hk), Q. Yang (qyang62-c@my.cityu.edu.hk): Department of Chemistry, City University of Hong Kong, Hong Kong, China

K. Zhai (kpzhai@semi.ac.cn), Y. Jin (jinya@semi.ac.cn), N. H. Zhu (nhzhu@semi.ac.cn): State Key Laboratory on Integrated Optoelectronics, Institute of Semiconductors, Chinese Academy of Sciences, Beijing, 100083, China

Z. Lin (21037392r@connect.polyu.hk): Department of Applied Physics and Research Institute for Smart Energy, The Hong Kong Polytechnic University, Hong Kong, China

Correspondence: E. Y. B. Pun and C. Wang

These authors contributed equally: S. Zhu, Y. Zhang, Y. Ren, Y. Wang

## 1. Simulated optical absorption results of the LNOI-2D material PDs.

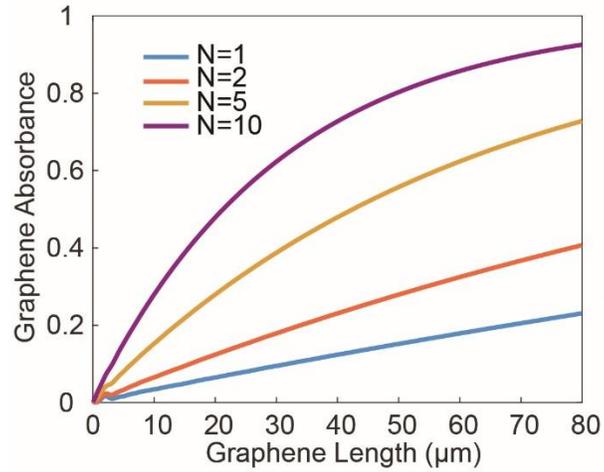

Fig. S1. The simulated optical absorption of the LNOI-graphene PD along with the graphene length under different layers. N refers to the number of graphene layers.

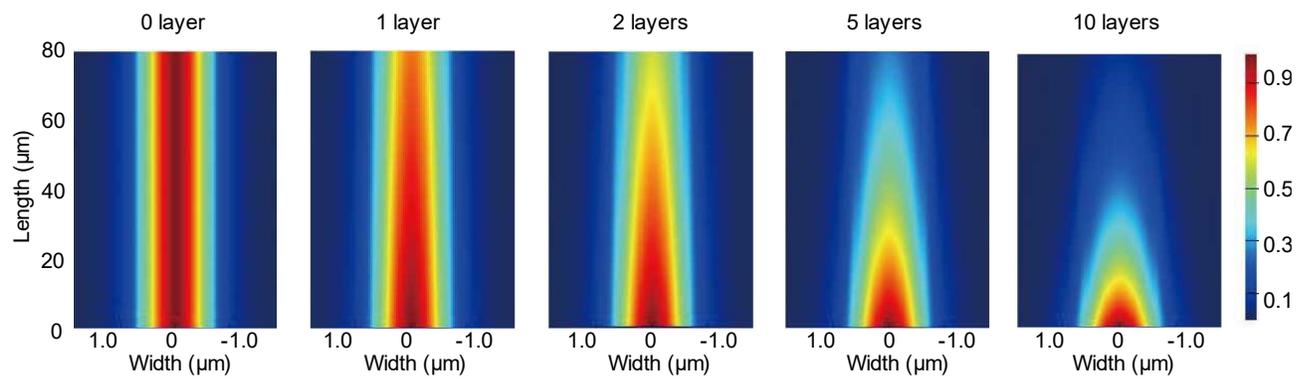

Fig. S2. The simulated electric field intensity evolution as a function of device length for different numbers of graphene layers, showing increased absorption for thicker graphene.

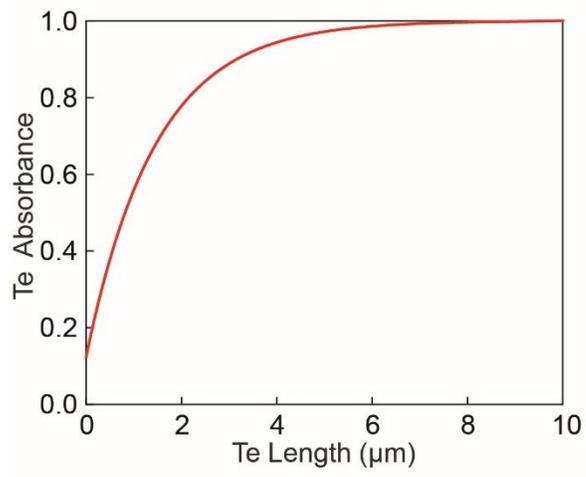

Fig. S3. The simulated optical absorption of the LNOI-Te PD along with the Te length.

## 2. TEM characterizations and Raman spectra of the 2D material for the LNOI-2D PDs

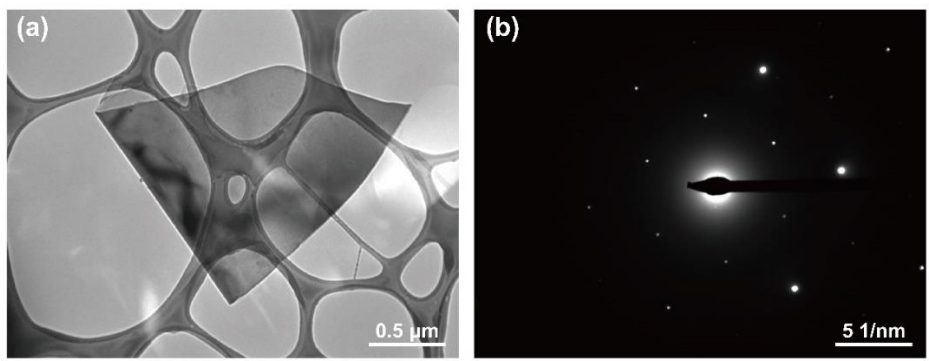

Fig. S4. (a) TEM image and (b) SAED of the multilayer graphene.

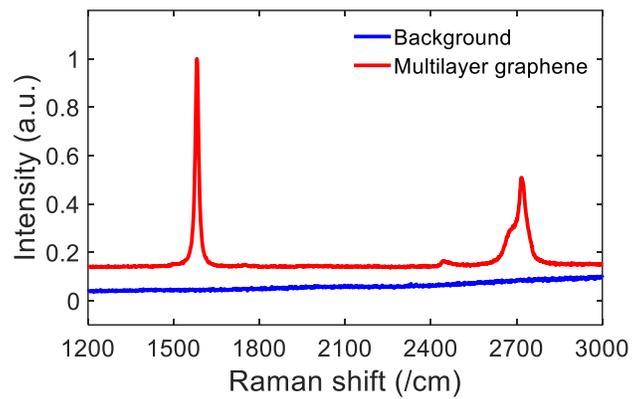

Fig. S5. Raman spectra of the multilayer graphene and bare substrate collected directly from the LNOI-graphene PD.

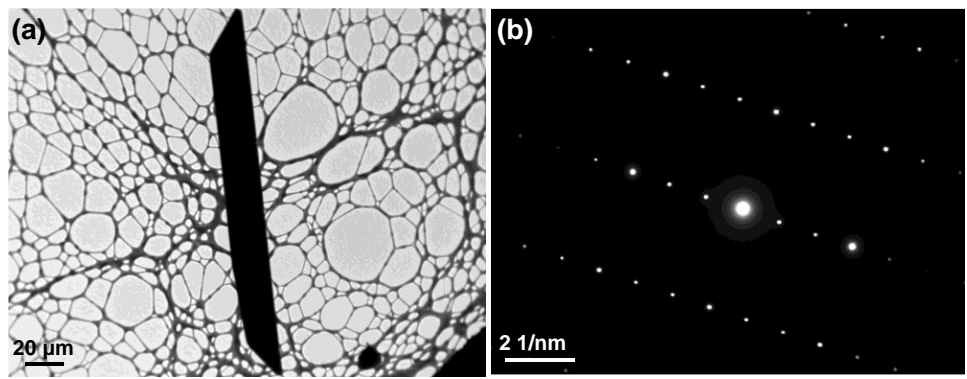

Fig. S6. TEM image (a) and SAED (b) of the Te.

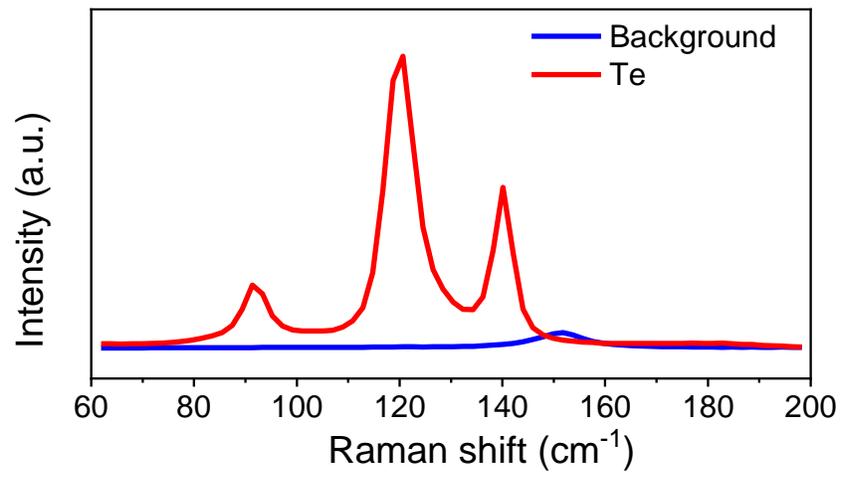

Fig. S7. Raman spectra of the Te and bare substrate collected directly from the LNOI-Te PD.

## 3. Experiment setups for the direct current and high-frequency response of the LNOI-2D material PD

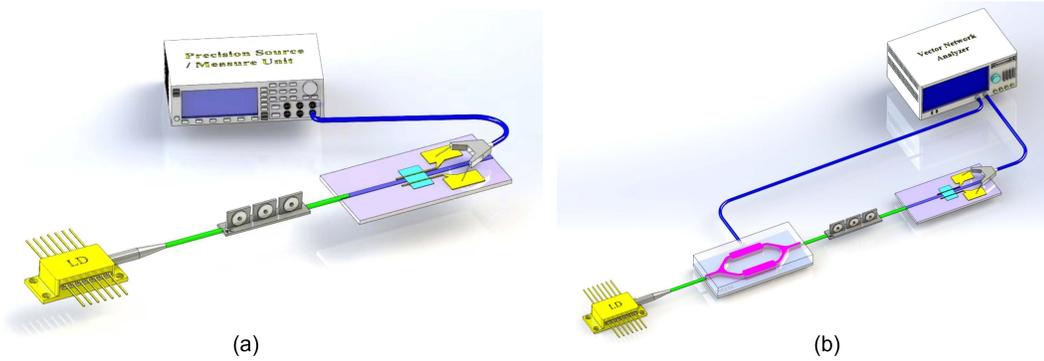

Fig S8. The experiment setups for the (a) direct current and (b) high-frequency responses of the fabricated LNOI-2D material PD.

## 4. Synthesis of Tellurium Nanoflakes

First, 1.5 g of polyvinylpyrrolidone (PVP) was dissolved in 16 mL of deionized water, then 46 mg of $Na_2TeO_3$ was added and dissolved into the PVP solution to form a clear solution. Add 1.66mL of ammonium hydroxide solution and 0.838mL of hydrazine monohydrate to the above solution in sequence. Transfer the prepared solution into a 25 mL Teflon-lined stainless steel autoclave and place it in the oven. Heat the autoclave from room temperature to 180 °C at a ramp rate of 3 °C/min and then maintain it at 180 °C for 30 h. After the reaction, the autoclave was taken out from the oven, and immediately cooled to room temperature with running water. The obtained product was purified and washed 3 times with deionized water, and centrifuged at 1000 rpm for 30 s for several times. The final product was redispersed in pure ethanol and transferred to the target substrate for characterization and device fabrication by drop casting.